\documentclass[twocolumn,aps,prb,showpacs,preprintnumbers,amsmath,amssymb,superscriptaddress]{revtex4-1}
\usepackage{graphicx}
\usepackage{bm}
\usepackage{gensymb}
\usepackage{color}

\usepackage{placeins}
\usepackage{soul}
\usepackage{url}

\begin{document}
\title{Dynamical vertex approximation in its parquet implementation: application to Hubbard nano-rings} 

\author{A.~Valli}
\affiliation{Institute for Solid State Physics, Vienna University of Technology, 1040 Wien, Austria}
\author{T.~Sch\"{a}fer}
\affiliation{Institute for Solid State Physics, Vienna University of Technology, 1040 Wien, Austria}
\author{P.~Thunstr\"{o}m}
\affiliation{Institute for Solid State Physics, Vienna University of Technology, 1040 Wien, Austria}
\author{G. Rohringer}
\affiliation{Institute for Solid State Physics, Vienna University of Technology, 1040 Wien, Austria}
\author{S.~Andergassen}
\affiliation{Faculty of Physics, University of Vienna, Boltzmanngasse 5, 1090 Vienna, Austria} 
\affiliation{Institut f\"{u}r Theoretische Physik and CQ Center for Collective Quantum Phenomena, Universit\"{a}t T\"{u}bingen, Auf der Morgenstelle 14, 72076 T\"{u}bingen, Germany} 
\author{G.~Sangiovanni}
\affiliation{Institute for Theoretical Physics and Astrophysics, University of W\"{u}rzburg, Am Hubland 97074 W\"{u}rzburg, Germany}
\author{K.~Held}
\affiliation{Institute for Solid State Physics, Vienna University of Technology, 1040 Wien, Austria}
\author{A.~Toschi}
\affiliation{Institute for Solid State Physics, Vienna University of Technology, 1040 Wien, Austria}



\date{\today}
\begin{abstract}
{We have implemented the dynamical vertex approximation (D$\Gamma$A) in its full parquet-based version 
to include spatial correlations on all length scales and in {\sl all} scattering channels. 
The algorithm is applied to study the electronic self-energies and the spectral properties 
of finite-size one-dimensional Hubbard models with periodic boundary conditions (nanoscopic Hubbard rings). 
From a methodological point of view, our calculations and their comparison to the results 
obtained within dynamical mean-field theory, plain parquet approximation, and the exact numerical solution, 
allow us to evaluate the performance of the D$\Gamma$A algorithm in the most challenging situation of low dimensions.
From a  physical perspective, our results unveil how  non-local correlations affect the spectral properties 
of nanoscopic systems of various sizes in different regimes of interaction strength.}
\end{abstract}
\maketitle

\section{Introduction}
Models and materials with reduced dimensionality typically  show enhanced correlation effects {\sl  beyond} the limit 
of standard density-functional or perturbation theory-based schemes, calling for corresponding developments of theoretical tools. 
From a general point of view, the challenge for a theoretical description is much bigger than in bulk systems: 
In three dimensions  (3D), even in the presence of strong electronic correlations, 
very accurate material calculations can be performed 
by means of the dynamical mean-field theory (DMFT),\cite{Metzner1989,Mueller-Hartmann1989,Georges1992,Georges1996} 
combined with {\sl ab-initio} methods.\cite{Anisimov1997,Lichtenstein1998,Kotliar2006} 
This is possible, because DMFT captures, non-perturbatively, 
the purely {\sl local} part of electronic correlations, which drives 
most important phenomena of correlated electrons in the bulk, such as, 
e.g., the Mott-Hubbard metal-insulator transition (MIT).\cite{Mott_MIT} 
Formally, DMFT becomes exact in the limit of infinite dimensions\cite{Metzner1989} 
where all non-local correlations in space are averaged out. 
Corrections to DMFT in finite-dimensional systems originate from non-local correlations. 
While in 3D deviations from the DMFT description become predominant
only in specific parameter regimes,\cite{Fuchs2011,Rohringer2011}
e.g., in the proximity of a second order phase transition,\cite{Rohringer2011} 
the situation is completely different in case of lower dimensions. 
In fact, reducing the dimensionality magnifies effects of {\sl non-local} correlations, 
undermining the main assumption of DMFT. 
Already for extended 2D systems, the physics deviates qualitatively from the DMFT predictions, 
e.g., the Mott-Hubbard MIT is found to disappear in a weak-coupling crossover in the 
phase-diagram of the 2D Hubbard model.\cite{Vilk1996,Borjesza2003,Schaefer-arXiv2014} 
Evidently,  even stronger non-local effects can be expected 
if the dimensionality is further reduced towards 1D or 0D. 

As for the theoretical description of electronic correlations at the
nanoscale, several algorithmic implementations based on DMFT have recently been implemented 
under the name of nano or real-space DMFT.\cite{Potthoff99,Florens08a,Snoek08,Valli2010,Titvinidze2012} 
Despite some technical differences, all these algorithms essentially extend the DMFT scheme to
finite-size and possibly non-translational invariant systems. 
The common idea consists in solving simultaneously several single impurity problems 
for calculating, separately, the local self-energies of the different sites composing the system of interest, 
while the DMFT self-consistency is then enforced at the level of the whole nanostructure. 
This way, a number of interesting results have been obtained both for model\cite{Florens08a,Valli2012,Rotter2013} 
and realistic studies.\cite{Jacob10,Das2011,Karolak11,Turkowski12}
However, the applicability of these DMFT-based methods is restricted 
to the weakly correlated and/or the high-temperature regime, where the effects of non-local correlations 
are weaker\cite{Katanin2009,Fuchs2011,Rohringer2011} and can be, to a certain extent, neglected. 
Such limitations were also openly discussed in the previous literature,\cite{Valli2010,Valli2012} 
where numerical comparisons between DMFT-based calculations and exact solutions (where available) 
have shown large deviations already in the intermediate coupling regime. 

A promising theoretical answer to this challenging situation has already been proposed, 
but not implemented, in Ref.\ \onlinecite{Valli2010}:
The application of diagrammatic extensions\cite{FLEXDMFT,Toschi2007,Held2008,DF,multiscale,1PI,DMF2RG} of DMFT such
as the dynamical vertex approximation (D$\Gamma$A)\cite{Toschi2007,Held2008} for nanoscopic systems (nano-D$\Gamma$A).  
The basic idea of D$\Gamma$A is the following: Instead of assuming the
locality of the one-particle self-energy [$\Sigma({\bf k},\omega) = \Sigma(\omega)$], as
in DMFT, one raises the assumption of the locality to a {\sl higher}
level of the diagrammatics, i.e., from the one- to the two-particle irreducible
vertex ($\Gamma_{i\rm rr}$)\cite{note_vertex,Rohringer2012,Schaefer2013} 
Once local vertex functions are computed, e.g., with the same impurity solvers used
for the standard DMFT,\cite{Toschi2007,Held2008,Hafermann2009,Kunes2011,Rohringer2012,Schaefer2013,Hafermann2014} 
non-local correlation effects can be directly included through diagrammatic relations, 
e.g., in the most general case, through the parquet equations.\cite{parquet} 

In the specific case of the D$\Gamma$A implementation for nanoscopic
systems,\cite{Valli2010} the nano-D$\Gamma$A algorithm requires a separate calculation of the local irreducible
vertex function for each inequivalent site of the nanostructure.
The inclusion of the non-local effects should be performed  at the
level of the whole nanostructure via a self-consistent solution of the parquet equations.  
This procedure is less demanding than the exact treatment of the
corresponding quantum Hamiltonian: the exponential scaling with the number of sites 
required for a diagonalization of the Hamiltonian,
is replaced by  a polynomial effort to solve the parquet equations.  
Moreover, the necessity of calculating the vertex functions only locally, mitigates secondary (but important)
numerical problems such as the sign-problem in quantum Monte Carlo (QMC) solvers. 
Nonetheless, the overall numerical efforts for treating the parquet equations has limited, so far, 
a wide application of D$\Gamma$A-based methods in their more complete (parquet-based) form: 
Hitherto, all successful applications of D$\Gamma$A to 2D and 3D systems have been performed in cases where
fluctuations in a given scattering channel predominate.\cite{Katanin2009,Rohringer2011,Schaefer-arXiv2014} 
In this case the solution of the parquet equation can be replaced by a much simpler ladder resummation 
performed in the most relevant channel only.\cite{Katanin2009} 
We note, in passing, that such considerations apply, with very few exceptions,\cite{Yang2011,DMF2RG} 
also to almost all other diagrammatic extensions of DMFT. 
For similar reasons, no application of the nano-D$\Gamma$A algorithm, 
as illustrated in Refs.\onlinecite{Valli2010} and \onlinecite{Valli2012} has been realized hitherto. 
Exploiting the constant improvements of the numerical
performance both in the DMFT calculations of vertex functions,\cite{Rohringer2012,Schaefer2013,Hafermann2014} as well
as in the numerical solution of the parquet equations,\cite{Yang2009,Tam2013,note_parquet} we will present here our first
results of the full (i.e., parquet-based) nano-D$\Gamma$A, applied to a set of correlated nanoscopic rings of increasing size. 

The importance of the results presented in the following is twofold, and goes beyond the demonstration of a full applicability
of the algorithm proposed in Ref.\ \onlinecite{Valli2010}:  Physically, our calculations allow to understand 
the interplay of local and non-local correlations in spectral and transport properties of finite systems of different sizes; 
from a methodological perspective, the application of a full (parquet-based) D$\Gamma$A scheme 
to these nanoscopic systems represent one of the most severe benchmarks conceivable for this theoretical approach. 
In fact, the accuracy of a D$\Gamma$A calculation depends on the correctness 
of the locality assumption for the two-particle irreducible vertex functions. 
Heuristically, this assumption looks plausible for 3D and 2D systems with local interactions, 
where strong spin, charge, and pair fluctuations are already generated 
by the corresponding collective modes built on local irreducible vertices.  
Numerically, a direct verification of the D$\Gamma$A assumption is difficult in 2D or 3D:
While the irreducible vertex surely displays a strong frequency dependence,\cite{Rohringer2012,Schaefer2013} 
taken into account by the D$\Gamma$A, its independence on momentum has been shown explicitly 
only in few calculations\cite{Maier2006} beyond DMFT, where the momentum-dependence was found to be weak. 
In this work, we focus instead on systems where an exact numerical solution is available, so that
both, the D$\Gamma$A performances and assumptions, can be tested. 
Let us emphasize that the low connectivity and the peculiarity 
of 1D physics represent the most challenging situation for D$\Gamma$A. 
In this perspective, our numerical analysis will also allow to draw conclusions, 
on a more quantitative ground, on the physical content of parquet-based approximations. 
The paper is organized as follows: In Sec.\ \ref{sec:model}, we introduce the
general properties of the nanoscopic systems under consideration, namely Hubbard-rings of
different sizes. In Sec. \ \ref{sec:method-DGA}, we discuss the parquet implementation of D$\Gamma$A. 
In Sec.\ \ref{sec:results}, we present the numerical parquet D$\Gamma$A results, 
while in Sec. \ \ref{sec:ladder} we also make a comparison with data obtained 
within the ladder approximation of the D$\Gamma$A scheme.  
Finally, Sec. \ \ref{sec:SEC} provides a summary and our conclusions, 
while the appendix contains the technical details of the numerical calculations. 

\begin{figure}[b!]
\includegraphics[width=0.5\textwidth, angle=0]{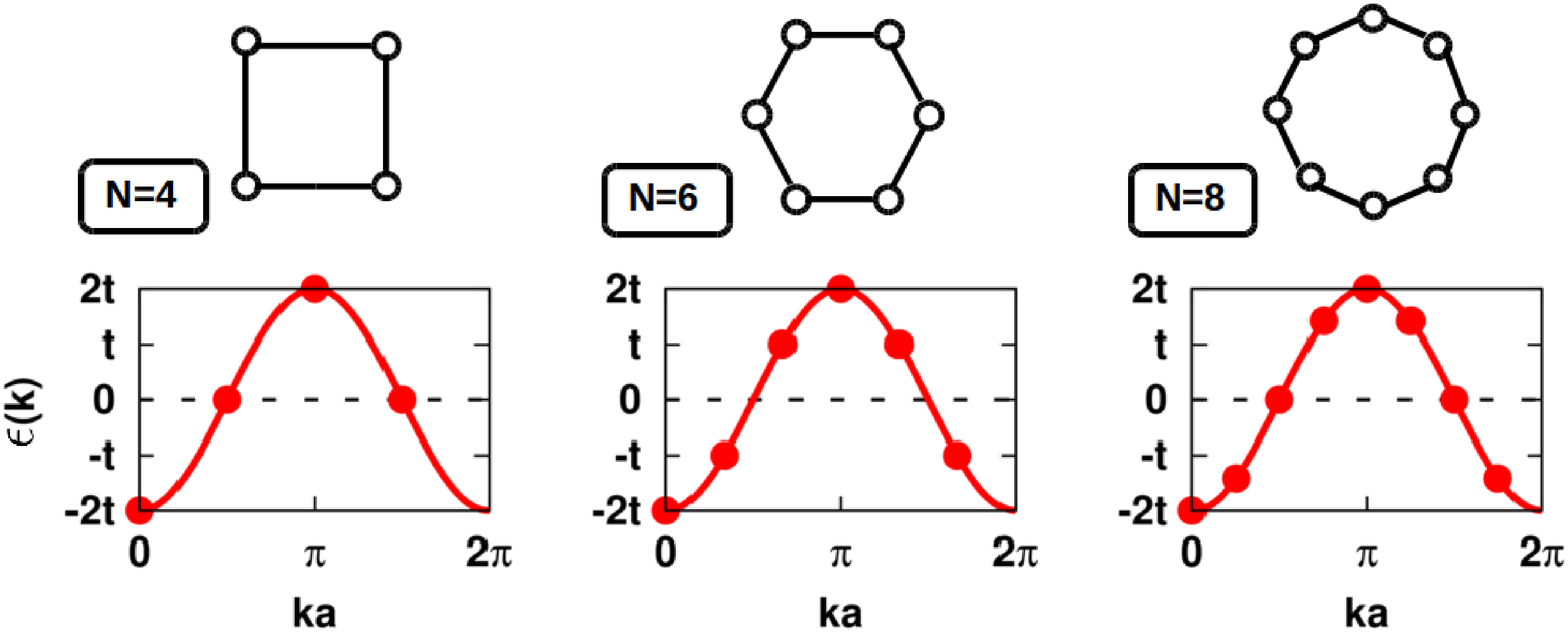}
\caption{(Color online) Energy-momentum dispersion relation $\epsilon(k)$ with  respect to the Fermi level $\mu$ (dashed line) 
for nano-rings with $N\!=\!4,6,8$ sites. The symbols denote the discrete eigenstates 
corresponding to the allowed values of the momentum: $k\!=\!2\pi n/N$, with $n\!\in\!\mathbb{N}$.}
\label{fig:epsk}
\end{figure} 

\section{Modelling the nano-rings}
\label{sec:model}
The correlated nanoscopic rings considered in the following consist of $N$ isolated correlated atoms, 
arranged in a chain with periodic boundary conditions, and described by the Hubbard Hamiltonian
\begin{equation}
 H = - t \sum_{\sigma} \sum_{i=1}^{N} \big( c^{\dagger}_{i\sigma} c^{\phantom{\dagger}}_{i+1\sigma} 
                                                                 + c^{\dagger}_{i+1\sigma} c^{\phantom{\dagger}}_{i\sigma} \big) 
       + U \sum_{i=1}^{N} n_{i\uparrow} n_{i\downarrow} 
\end{equation}
where $c^{\dagger}_{i\sigma}$ ($c^{\phantom{\dagger}}_{i\sigma}$) 
denote the creation (annihilation) operators of an electron on site $i$ with spin $\sigma$, 
fulfilling the periodic boundary conditions $c_{(i+N)\sigma} \!=\! c_{i\sigma}$, while 
$n_{i\sigma}\!=\!c^{\dagger}_{i\sigma}c^{\phantom{\dagger}}_{i\sigma}$ denotes the number operator; 
the parameters $t$ and $U$ denote the nearest-neighbor (NN) hopping amplitude and the on-site Hubbard interaction, respectively. 
Due to the translational invariance of the system, granted by the periodic boundary conditions of the ring, 
it is convenient to formulate the hopping term in the reciprocal space, 
yielding a tight-binding dispersion $\epsilon(k)\!=\!-2 t \ {\rm cos}(ka)\!-\!\mu$, where $\mu$ is the chemical potential. 
In the following, we set the lattice spacing $a\!=\!1$ and consider rings with $N\!=\!4,6,8$ sites. 
We restrict ourselves to the half-filled case, i.e., $\mu\!=\!U/2$, 
where electronic correlations stemming from the local Hubbard interaction are expected to be most effective. 
Under these conditions, all rings display a particle-hole symmetric density of states, 
and in particular, in the non-interacting case ($U\!=\!0$) the systems 
display either a ''band'' gap (as in the case of the $N\!=\!6$ sites ring) 
or a 2-fold degenerate state at the Fermi level (as in the case of $N\!=\!4,8$ sites rings). 
The rings  considered in this work and the corresponding dispersions $\epsilon(k)$ are shown 
in the upper and lower panel of Fig.\ \ref{fig:epsk}, respectively.

\section{Parquet-based implementation of the nano-D$\Gamma$A} 
\label{sec:method-DGA}
We recall that the idea of D$\Gamma$A is to apply the locality assumption of DMFT 
at an higher level of the diagrammatics: While in DMFT all one-particle irreducible (1PI) one-particle diagrams 
(i.e., the self-energy $\Sigma$) are assumed to be purely local, 
D$\Gamma$A confines the locality to the two-particle irreducible (2PI) two-particle diagrams, 
i.e., the  fully irreducible vertex $\Gamma_{\rm irr} $ is approximated by all local Feynman diagrams.\cite{note_diagrams}
Hence, in the D$\Gamma$A framework, the purely local, but frequency-dependent,\cite{note_freqs}
2PI vertex $\Gamma_{\rm irr} =\Lambda_{iiii}^{\omega\nu\nu'}$ is calculated 
for a site $i$ and then used as the input for the parquet equations. 
In practice, this vertex is obtained by solving the Anderson impurity model (AIM) numerically. 
Hence, non-local correlations on top of the DMFT solution are generated in {\sl all} scattering channels 
by the (numerical) solution of the parquet equations,\cite{parquet}
without any restriction to specific (ladder) subsets of diagrams.\cite{Katanin2009} 
For the sake of clarity, we should emphasize here that this is different 
from the so-called parquet approximation (PA). In fact, the PA corresponds to approximating the 2PI vertex 
with the bare interaction of the theory (e.g., $\Gamma_{\rm irr}\!=\!U$) in a merely perturbative fashion. 
On the contrary, in D$\Gamma$A all non-perturbative DMFT correlations, 
which control, e.g., the physics of the Mott-Hubbard transition, 
are actually  included through the frequency dependent $\Gamma_{\rm irr}\!=\!\Lambda^{\omega\nu\nu'}$, 
and non-local correlations beyond DMFT are generated via the solution of the parquet equations. 

\begin{figure}
\includegraphics[width=0.27\textwidth, angle=0]{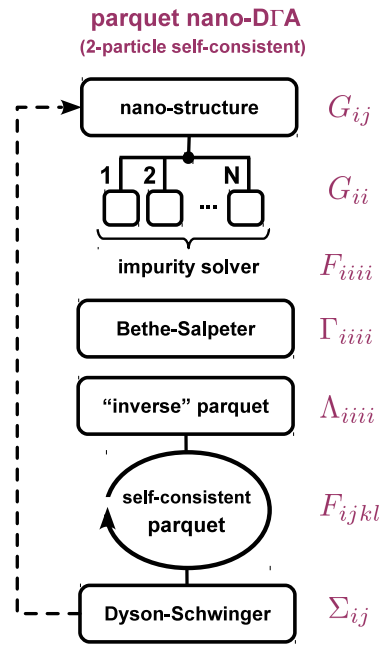}
\caption{(Color online) Flowchart of the parquet implementation of the nano-D$\Gamma$A. See text for a related discussion. }
\label{fig:flowchart}
\end{figure} 

The specific implementation of the parquet-based D$\Gamma$A scheme 
for the case of nanoscopic systems, such as the Hubbard nano-rings, 
is briefly sketched in the flowchart of Fig.\ \ref{fig:flowchart}, 
and incorporates all main aspects of the original proposal of Ref.\ \onlinecite{Valli2010}. 
Let us start by recalling the DMFT scheme for a nanoscopic system with $N$ constituents (e.g., atoms), 
which is self-consistent at the one-particle level only. 
The first step consists in mapping the full problem onto a set of auxiliary AIMs, one for each of the $N$ sites of the nanostructure. 
Each auxiliary problem is characterized by a dynamical Weiss field (i.e., the non-interacting Green's function of the AIM) ${\cal G}_{0i}(\nu)$. 
The numerical solution of the AIM  yields the local Green's function $G_{ii}(\nu)$ 
and the local DMFT self-energy $\Sigma_{ii}(\nu)={\cal G}_{0,ii}^{-1}(\nu) \!-\! G_{ii}^{-1}(\nu)$. 
Through the Dyson equation the local (yet site-dependent) DMFT self-energies 
determine the new non-local Green's function $G_{ij}$, 
and the self-consistency is realized at the level of the whole nanostructure. \\
In the case of the D$\Gamma$A this procedure is raised to the two-particle level.
For each inequivalent AIM, the local 2PI vertex function is computed as following 
(for the sake of clarity, we omit the temporal and spatial indexes in this derivation, 
and yet recall that those steps are performed at the local level of each AIM). 
Typically, one first calculates the generalized local susceptibility $\chi$ 
with the impurity solver of the AIM.\cite{Toschi2007,Rohringer2012} 
Then one extracts the full vertex $F$ from the local susceptibility as
\begin{equation} 
\chi = \chi_0 - \frac{1}{\beta^2} \chi_0 F \chi_0, 
\end{equation}
where $\chi_0$ is the bubble part of $\chi$, while $F$ includes all possible vertex corrections. 
In order to obtain the fully irreducible vertex $\Lambda$, it is necessary to separate the two-particle
reducible ($\Phi_r$) and irreducible ($\Gamma_r$) contributions to the full vertex $F$ in each scattering channel $r$, 
by solving the corresponding Bethe-Salpeter equation 
\begin{equation} \label{eq:BSE}
F = \Gamma_r + \Phi_r = \Gamma_r + \int \Gamma_r GG F ,
\end{equation}
where the integral symbol denotes a summation 
over all internal degrees of freedom (e.g., frequencies, spin, ...). 
The fully irreducible vertex $\Lambda$ is obtained from $F$ and the $\Phi_r$'s 
by inverting the parquet equation of the AIM 
\begin{equation} \label{eq:parquet}
F = \Lambda + \sum_r \Phi_r .
\end{equation}
Further details can be found in Ref.\ \onlinecite{Rohringer2012}, which provides a comprehensive discussion 
of the local two-particle vertex functions and of the parquet equations in a unified formalism.  \\
Once all inequivalent local 2PI vertices $\Lambda_{iiii}$ are obtained for each site $i$, 
they are used as an input for the solution of the parquet equations for the whole nanoscopic system. 
This yields the non-local full two-particle vertex function $F_{ijkl}$ 
and, through the Dyson-Schwinger equation, the non-local self-energy 
\begin{equation}  \label{eq:DSE}
\Sigma_{ij} = \frac{Un}{2} \delta_{ij} - \frac{U}{\beta^2} \int \sum_{klm} G_{ik} G_{il} G_{im} F_{klmj}^{\uparrow\downarrow} ,
\end{equation}
where the integral symbol, as above, denotes a summation over all the internal degrees of freedom, 
while the sum over the spatial indices of the nanoscopic system is explicit. 
For clarity we recall that the local Hartree shift of the self-energy $\Sigma_{\rm H}\!=\!Un/2$ 
is already included in the definition of the chemical potential. 
The set of equations (\ref{eq:BSE}-\ref{eq:DSE}) can be solved self-consistently 
until the non-local self-energy (\ref{eq:DSE}) is converged. \cite{Yang2009,Tam2013} 
The flowchart of the parquet D$\Gamma$A  is shown schematically in Fig.\ \ref{fig:flowchart}. 
Finally, after having determined $\Sigma_{ij}$ one can either skip the outermost loop, 
i.e., updating the AIM and simply start from ${\cal G}_{ii}$ of DMFT, as we did in the present paper, 
or one can perform fully self-consistent calculations;
In the latter case the ${\cal G}_{ii}$  of the corresponding inequivalent AIMs has to be adjusted 
to yield the given D$\Gamma$A ${G}_{ii}$ from the previous iteration before recalculating the 2PI vertex 
(which is defined diagrammatically in terms of $U$ and ${G}_{ii}$). One then needs to iterate this scheme until convergence. 
We refer to Refs.\ \onlinecite{RohringerPhD}  and \onlinecite{DGAJuelich} for a more detailed discussion of the D$\Gamma$A scheme, 
including also the Feynman diagrams and all the equations. 
From the flowchart of the algorithm, one can clearly see how in the D$\Gamma$A non-local correlations beyond DMFT 
are systematically generated in {\sl all} scattering channels in a two-particle self-consistent framework. \\

\section{Results}
\label{sec:results}
In the following we present the numerical results for all Hubbard nano-rings discussed in Sec.\ \ref{sec:model}, 
characterized by the dispersions $\epsilon(k)$ shown in Fig.\ \ref{fig:epsk} (lower panels). 
For each system we compare different approximations, 
i.e., DMFT, PA, and parquet D$\Gamma$A, to the exact QMC solution. 
Each method employed in this work is associated to a specific diagrammatic content, 
as discussed in Sec.\ \ref{sec:method-DGA}, which allows us to understand the relevance 
of specific subsets of Feynman diagrams for the description of the systems considered. 
Later, in Sec. \ \ref{sec:ladder}, we will also compare the self-energies shown in the following with the ones 
obtained within the ladder approximation of the D$\Gamma$A scheme, which represents the typical framework 
of previous D$\Gamma$A calculations \cite{Toschi2007,Katanin2009,Rohringer2011,Schaefer-arXiv2014} \\

We will discuss the results obtained for the electronic self-energy $\Sigma(k,\imath\nu_n)$, the local Green's function $G_{ii}(\tau)$, 
and the two-particle irreducible local (i.e., DMFT) vertex function $\Lambda_{iiii}^{\omega\nu\nu'}$. 
The analysis of the self-energy allows to resolve a $k$-selective behavior in the (discrete) reciprocal space. 
In particular, we analyze two low-energy parameters, i.e., 
the scattering rate $\gamma(k)\!\equiv\!-2 {\rm Im}\Sigma(k,\imath\nu_n\!\rightarrow\!0)$, 
which corresponds to a damping or to the inverse lifetime of quasi-particle excitations in the Fermi liquid regime, 
and the (static) renormalization of the bare dispersion $\Delta(k)\!\equiv\!{\rm Re}\Sigma(k,\imath\nu_n\!\rightarrow\!0)$. 
We will discuss the effect of the local and non-local self-energy on the low-energy spectral properties of the system 
which can be deduced by the local Green's function, 
and is related to the $k$-resolved spectral function $A(k,\nu)$ by 
\begin{equation}
G_{ii}(\tau) \!=\! \sum_k \int_{-\infty}^{\infty} \! d\nu \frac{{\rm e}^{-\tau\nu}}{1+{\rm e}^{\beta\nu}} A(k,\nu) .
\end{equation}
The value of the Green's function at $\tau\!=\!\beta/2$ represents an estimate of the 
value of the local spectral function at the Fermi level 
(averaged over an energy window proportional to the temperature $T$), i.e., 
\begin{equation}
-\beta G_{ii}(\beta/2) \approx \pi \sum_k A(k,0). 
\end{equation}  
In order to understand the non-local self-energy corrections beyond mean-field, 
we will also relate our results to the frequency structure of the local 2PI vertex ($\Lambda_{iiii}$), 
which is the input for the parquet equations of the D$\Gamma$A. 
To this end, the generalized susceptibility of the AIM is computed 
and the 2PI vertex is obtained following the steps discussed in Sec.\ \ref{sec:method-DGA}. 
For the analysis of the 2PI vertex we will adopt the notation of Refs.\ \onlinecite{Rohringer2012} and \onlinecite{RohringerPhD}, 
and in particular we will consider the 2PI vertex in the (particle-hole) density and magnetic channels\cite{note_channels} 
with respect to their static asymptotics, i.e., $\Lambda_{d,m}^{\omega\nu\nu'} \mp U$.  
Moreover, the comparison to numerically exact two-particle vertex functions, and in particular the fully irreducible one, 
will also allow us to directly test the assumption behind the D$\Gamma$A, i.e., the locality of $\Lambda$. 

In the following, we will start presenting in Sec.\ \ref{sec:results-nc6} the numerical results 
for the $N\!=\!6$ sites nano-ring before turning, in Sec.\ \ref{sec:results-nc4,8}, to  nano-rings with $N\!=\!4,8$ sites. 
The reason for this choice is that the low-energy physics of the $N\!=\!6$ sites ring 
is controlled by an energy scale $\Delta_0\!=\!2t$, associated to the gap in the non-interacting density of states, 
which makes the system behave more similarly to a correlated band insulator. 
On the other hand, the $N\!=\!4,8$ sites rings are both characterized by  the presence of a 2-fold degenerate state 
at the Fermi energy, which induces a physical behavior similar to the one of a correlated metal.

\begin{figure}[b!]
\includegraphics[width=0.47\textwidth, angle=0]{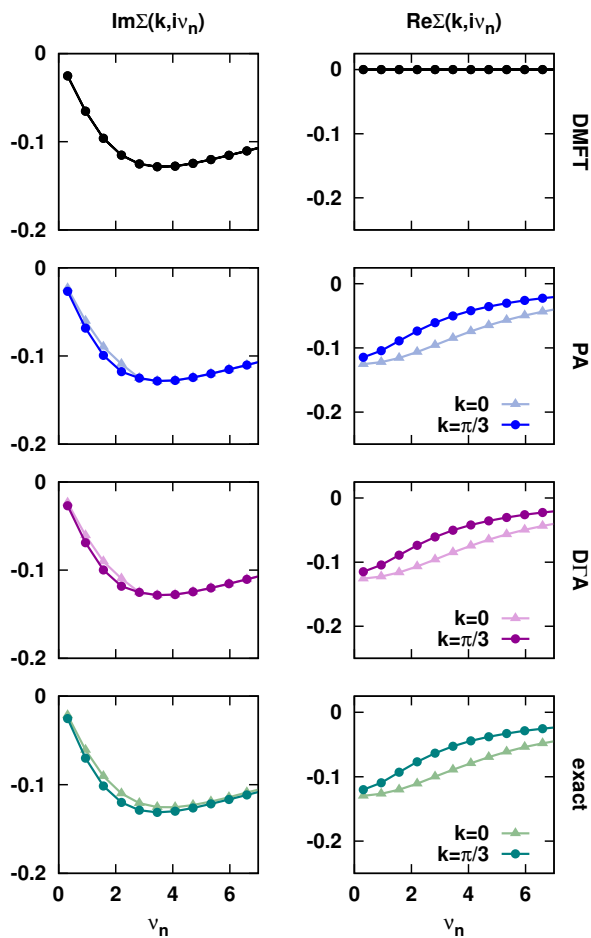}
\caption{(Color online) Comparison between the local DMFT self-energy in Matsubara representation 
and the $k$-resolved PA, D$\Gamma$A, and the exact self-energy, for representative $k$-points in the DBZ. 
In this case, including the full frequency dependence of $\Lambda$ results in negligible corrections 
to the static PA. Parameters: $N\!=\!6$, $U\!=\!2t$ and $T\!=\!0.1t$.}
\label{fig:siwk_nc6U2T0.1-parquet}
\end{figure}

\subsection{$N\!=\!6$: ''insulating'' ring}
\label{sec:results-nc6}
In previous works,\cite{Valli2010,Valli2012} we analyzed by means of nano-DMFT the electronic and transport properties 
in a $N\!=\!6$ Hubbard nano-ring in the presence of hybridization with a substrate. 
In particular, it was shown\cite{Valli2010,Valli2012} that in the weak-hybridization regime, 
and especially in the case of an isolated nanostructure considered here, 
non-local correlations beyond DMFT are not negligible 
and have an important effect on the electronic and transport properties of the system. 
In fact, local electronic correlations within DMFT shrink the gap with respect to 
the value predicted, e.g, within a H\"{u}ckel\cite{Hueckel} picture, 
akin to what happens in bulk correlated band insulators.\cite{Kunes2008,Sentef2009} 
Instead, the numerically exact solution (obtained by means of Hirsch-Fye\cite{Hirsch86a} QMC simulations, see Appendix for details) 
shows that non-local correlations yield a wider spectral gap 
due to the effective renormalization of the hopping parameter by a non-negligible NN self-energy in real space.  
With increasing hybridization between the correlated sites and the substrate, 
non-local spatial correlations are gradually suppressed, while local correlations remain sizeable. \cite{Valli2012} \\
We now extend the analysis done in previous works including non-local correlations via the parquet D$\Gamma$A,  
which yields a qualitative and quantitative agreement with exact QMC simulations. 
We note that the D$\Gamma$A results for the $N\!=\!6$ sites ring presented in this section, 
are expected to hold also for generic semiconducting nano structures in the weak- and intermediate-coupling regime, 
i.e., where the bare interaction $U$ is comparable with the size of the gap. 
We have tested this claim to hold for another gapped Hubbard ring with $N\!=\!10$ sites 
and we have found a similar agreement (not shown). 

In Fig.\ \ref{fig:siwk_nc6U2T0.1-parquet} we compare the local DMFT self-energy with the $k$-resolved self-energy 
for representative $k$ points\cite{note_siwk} in the discrete Brillouin zone (DBZ), namely $k\!=\!0$ and $k\!=\!\pi/3$, 
obtained by means of PA, D$\Gamma$A and exact QMC solution. 
Concerning the imaginary part of the self-energy ${\rm Im}\Sigma(k,\imath\nu_n)$, 
one can note that all the approximations employed provide a qualitative and \emph{quantitative} agreement with the exact solution. 
The system displays a low scattering rate $\gamma_k$, which is consistent with the picture of an insulating ground state 
reminiscent of the band gap of the non-interacting spectral function (renormalized by electronic correlations), 
rather than driven by a Mott MIT. 
More specifically, the exact QMC self-energy displays a weak $k$-dependence at low frequencies, 
resulting in a slightly different scattering rate $\gamma_k$ for different $k$ points in the DBZ. 
While this feature cannot be reproduced within DMFT by definition, 
it is well captured including non-local correlations beyond mean-field. 
Concerning the real part of the self-energy, we observe that 
within DMFT ${\rm Re}\Sigma(\imath\nu_n)\!=\!0$, i.e., all contributions averaging out in the local picture, 
except for the Hartree term, which is included in the redefinition of the chemical potential, 
i.e., $\mu \!\rightarrow\! \mu\!-\!U/2$ at half-filling. 
On the contrary,  including non-local correlations beyond mean-field 
we find a sizeable $k$-dependent self-energy ${\rm Re}\Sigma(k,\imath\nu_n)$. 
In all cases the exact self-energy is quantitatively well reproduced.

\begin{figure}[b!]
\includegraphics[width=0.42\textwidth, angle=0]{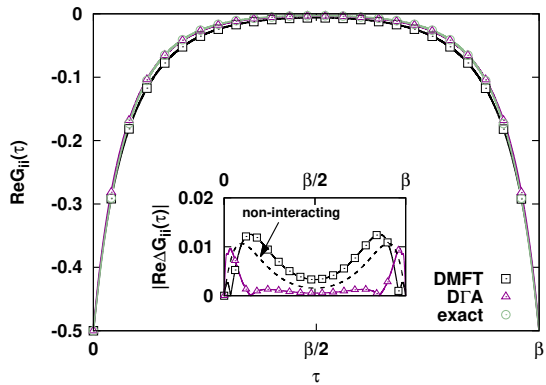}
\caption{(Color online) Comparison of the local Green's function $G_{ii}(\tau)$ obtained from the corresponding 
DMFT, D$\Gamma$A and exact self-energy. The inset shows the difference $\Delta G_{ii}(\tau)$ 
between the corresponding approximation and the exact solution. Parameters: $N\!=\!6$, $U\!=\!2t$ and $T\!=\!0.1t$.}
\label{fig:gtau_nc6U2T0.1}
\end{figure}

\begin{figure}[t!]
\includegraphics[width=0.5\textwidth, angle=0]{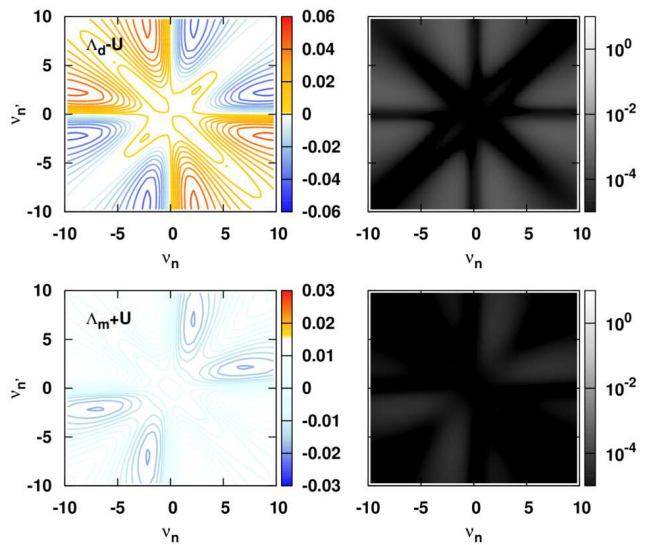}
\caption{(Color online) Local two-particle fully irreducible vertex calculated in DMFT 
in the (particle-hole) density and magnetic channels with respect to the static asymptotics, 
i.e.: $\Lambda_d\!-\!U$ (upper row) and $\Lambda_m\!+\!U$ (lower row), 
as a function of the two fermionic frequencies $\nu_n$ and $\nu_{n'}$, for bosonic frequency $\omega=0$.
The isoline plot (left panels) highlights the frequency and sign structure of the vertex, 
while the gray-scale density plot (right panels) shows its logarithmic intensity.  
Parameters: $N\!=\!6$, $U\!=\!2t$, and $T\!=\!0.1t$. }
\label{fig:fir_nc6U2T0.1} 
\end{figure}

Fig.\ \ref{fig:gtau_nc6U2T0.1} shows the effect of non-local correlations
on the local Green's function $G_{ii}(\tau)$. In the case of the $N\!=\!6$ sites ring, the interpretation of the results is straightforward. 
In fact, all methods predict an insulating solution, and this is reflected by $G_{ii}(\beta/2) \!\approx\! 0$ 
(which is a measure for the spectral weight in an energy interval $\sim\!T$ around the chemical potential; 
it is exactly zero only in the limit $T\!\rightarrow \!0$). 
However, at a closer look one can notice that the DMFT predicts more spectral weight $A(0)$, 
or equivalently a smaller value of the spectral gap, than the other methods.  
It is interesting to notice that, considering specifically $A(0)$, the DMFT is {\sl worse} than the non-interacting case, 
while obviously DMFT is superior in several other respects, e.g., lifetime of one-particle excitation. 
This is clearly shown in the inset of Fig.\ \ref{fig:gtau_nc6U2T0.1}, where we plot the difference $\Delta G_{ii}(\tau)$ 
of the local Green's function for the different approximations with respect to the one of the exact solution. 
Hence, we can ``disentangle'' the roles played by local and non-local correlations on an insulator considering that, in an insulator: 
i) taking into account only local correlations within DMFT reduces the non-interacting spectral gap,\cite{Sentef2009} and
ii) the non-local correlations in the exact solution display the opposite trend, as correctly described by the D$\Gamma$A. 
Indeed, the analytic continuation of the Green's function by means of the maximum entropy method (not shown) 
confirms the expectations, yielding a spectral gap $\Delta \approx 1.9t$ within DMFT 
and $\Delta \approx 2.2t$ within D$\Gamma$A and the exact solution, 
to be compared to the non-interacting value $\Delta_0 \!=\! 2t$.

We can understand the results obtained for both the self-energy and the local Green's function 
within the different approximations, by taking a closer look at the local fully irreducible vertex calculated from the  DMFT Green's function. 
The isolines and the density plot in the left and right panels of Fig.\ \ref{fig:fir_nc6U2T0.1}, respectively, 
highlight the sign and the logarithmic intensity of the frequency structure of $\Lambda_{d,m}$.  
The fully irreducible vertex displays the typical \emph{butterfly} structure previously reported,\cite{Rohringer2012,Schaefer2013} 
with positive and negative lobes decaying to the bare interaction value at high frequency (beyond the frequency range shown here). 
The frequency structure of the 2PI vertex beyond the static asymptotics is negligible with respect to the bare interaction $U\!=\!2t$. 
This is a consequence of the spectral gap, resulting in insulating Green's functions already within DMFT. 
The inversion of sign at low frequencies in the first and third quarter of the $(\nu,\nu')$ plane originates, instead, 
from the precursor lines of the Mott transition recently found in the DMFT phase diagram.\cite{Schaefer2013} 
The negligible frequency structure of the local 2PI vertex explains 
why the D$\Gamma$A self-energy does not deviate from the plain PA result in this case. 
On the other hand, the quantitative agreement with the exact QMC solution, 
suggests that the local D$\Gamma$A assumption of the 2PI is justified in this system. 
The direct numerical analysis of the exact 2PI vertex confirms that, 
besides a structure in momentum space, 
its overall values yield moderate corrections to the bare interaction $U\!=\!2t$ (not shown). 

\subsection{$N\!=\!4$ and $N\!=\!8$: ''correlated metallic'' rings}
\label{sec:results-nc4,8}
In contrast to the previous system, both the $N\!=\!4$ and $N\!=\!8$ sites rings 
are characterized by the presence of a (doubly degenerate) eigenstate 
at the Fermi level of the non-interacting density of states. 
For this reason we would expect them to display a similar behavior, 
and a different low-energy physics with respect to the $N\!=\!6$ sites ring. 
As we will show, this is only partially true. 

\begin{figure}[b!]
\includegraphics[width=0.47\textwidth, angle=0]{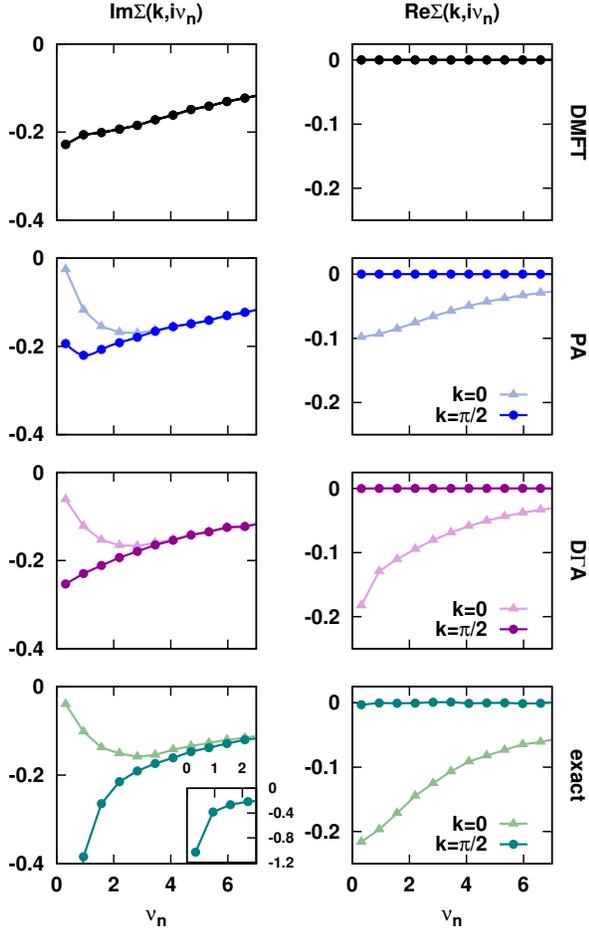}
\caption{(Color online) As in Fig.\ \ref{fig:siwk_nc6U2T0.1-parquet} but for the $N\!=\!4$ ring. 
In contrast to the previous case, including the full frequency dependence of $\Lambda$ 
leads to a substantial improvement of the D$\Gamma$A over the static PA. 
The inset shows the low-energy tendency toward a divergency of the exact self-energy for $k\!=\!\pi/2$. }
\label{fig:siwk_nc4U2T0.1-parquet}
\end{figure}

Let us start discussing the $k$-resolved self-energy of the $N\!=\!4$ ring, 
shown in Fig.\ \ref{fig:siwk_nc4U2T0.1-parquet} for representative $k$ points in the DBZ, 
namely $k\!=\!0$ and $k\!=\!\pi/2$ (the latter at the Fermi surface). 
In this case the DMFT self-energy displays a non-Fermi liquid behavior, 
characterized by a large yet finite scattering rate $\gamma$ (obviously independent on $k$). 
As we will see below, the system is not gapped in DMFT. 
The DMFT picture, however, is substantially changed by non-local correlations, as reflected in a strong $k$-dependent behavior 
of the self-energy, found within all approximations considered. 
In particular, away from the Fermi surface (e.g., at $k\!=\!0$) all approximations yield a low scattering rate $\gamma_{k\!=\!0}$ 
due to the bending towards zero of ${\rm Im}\Sigma(k,\nu_n)$.  
The situation is drastically different at the Fermi surface (e.g., at $k\!=\!\pi/2$), where in the exact solution, 
the divergent tendency of the self-energy marks the opening of a gap in the spectral function.  
Taking into account all scatting channels within the parquet  D$\Gamma$A formalism 
leads to an improvement with respect to the DMFT results. 
While the PA yields a sizeable scattering rate $\gamma_{k\!=\!\pi/2}\!\approx\!0.4$, 
including the frequency dependence of the fully irreducible vertex within D$\Gamma$A 
further enhances $\gamma_{k\!=\!\pi/2}$ and reproduce correctly the qualitative trend of the exact self-energy, 
as well as an overall better description of the ${\rm Re}\Sigma(k,\nu_n)$ with respect to PA and DMFT. 
The quantitative difference between the parquet D$\Gamma$A and the exact solution may originate 
either from the momentum dependence of the 2PI vertex, neglected in D$\Gamma$A, or by the lack of self-consistency. \\
Further insights can be obtained by considering the spin propagator 
$\chi_s^{\omega}(q)$, in particular at $\omega\!=\!0$. 
Within DMFT, we find that $\chi_s(q\!=\!\pi) \!<\! 0$. The unphysical value of the susceptibility indicates that 
the system is below the N\'{e}el temperature of DMFT, i.e., $T\!<\!T^{\rm DMFT}_{N}$, 
while no ordering is expected at finite temperature. 
Including non-local spatial correlations within the parquet D$\Gamma$A scheme reduces\cite{Vlik1997} $T_N$. 
However, it is plausible that the local physics described by DMFT, and hence the information 
encoded into the 2PI vertex of DMFT, can be very different from the local physics of the exact solution. 

\begin{figure}[t!]
\includegraphics[width=0.42\textwidth, angle=0]{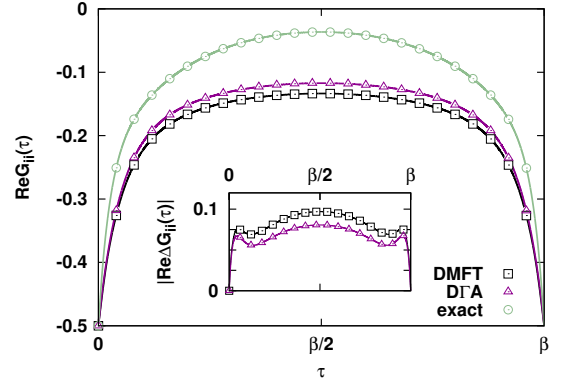}
\caption{(Color online) As in Fig.\ \ref{fig:gtau_nc6U2T0.1} but for the $N\!=\!4$ ring.}
\label{fig:gtau_nc4U2T0.1}
\end{figure}

\begin{figure}[t!]
\includegraphics[width=0.5\textwidth, angle=0]{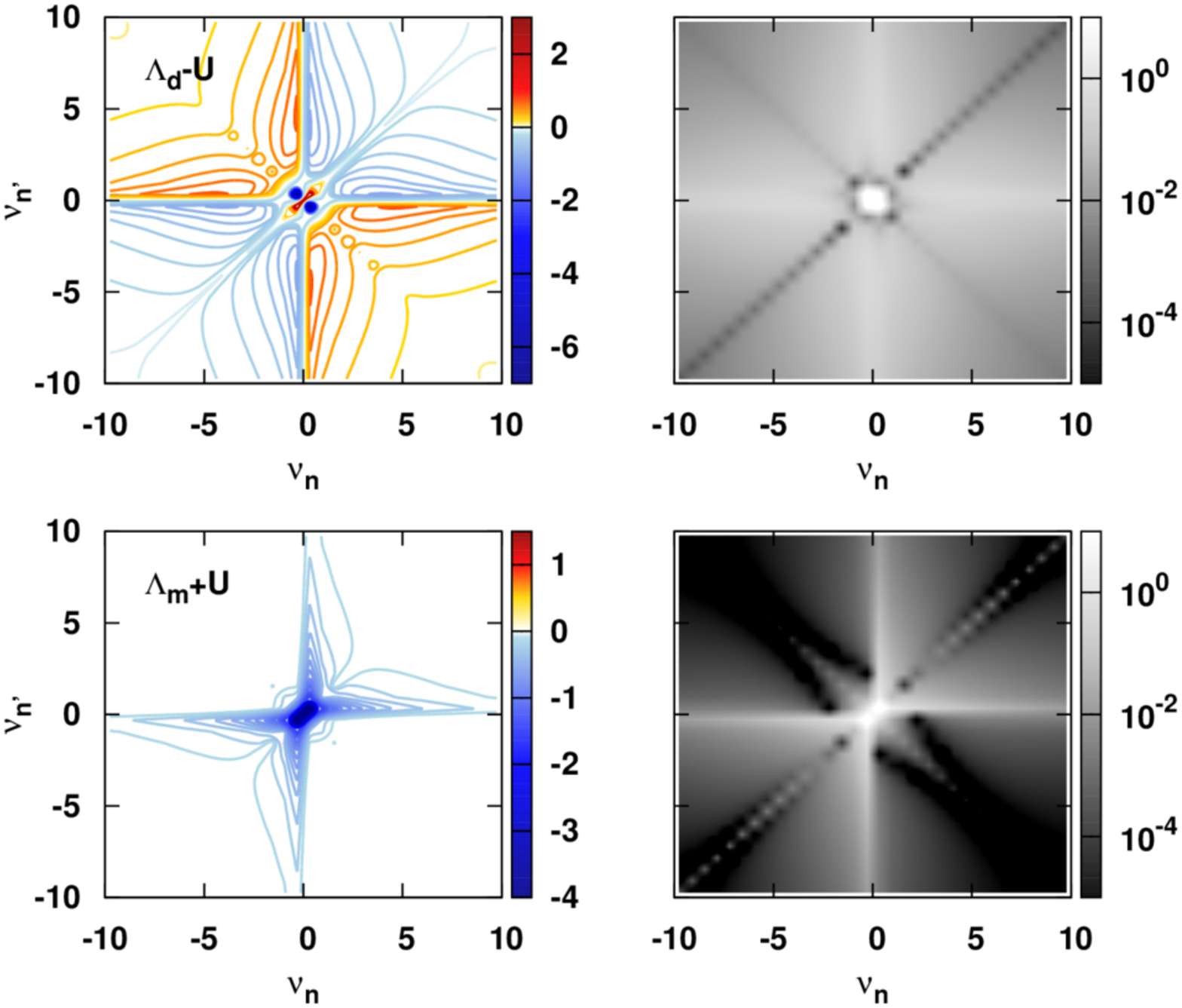}
\caption{(Color online) As in Fig.\ \ref{fig:fir_nc6U2T0.1} but for the $N\!=\!4$ ring.}
\label{fig:fir_nc4U2T0.1}
\end{figure}

\begin{figure*}
\includegraphics[width=1.0\textwidth, angle=0]{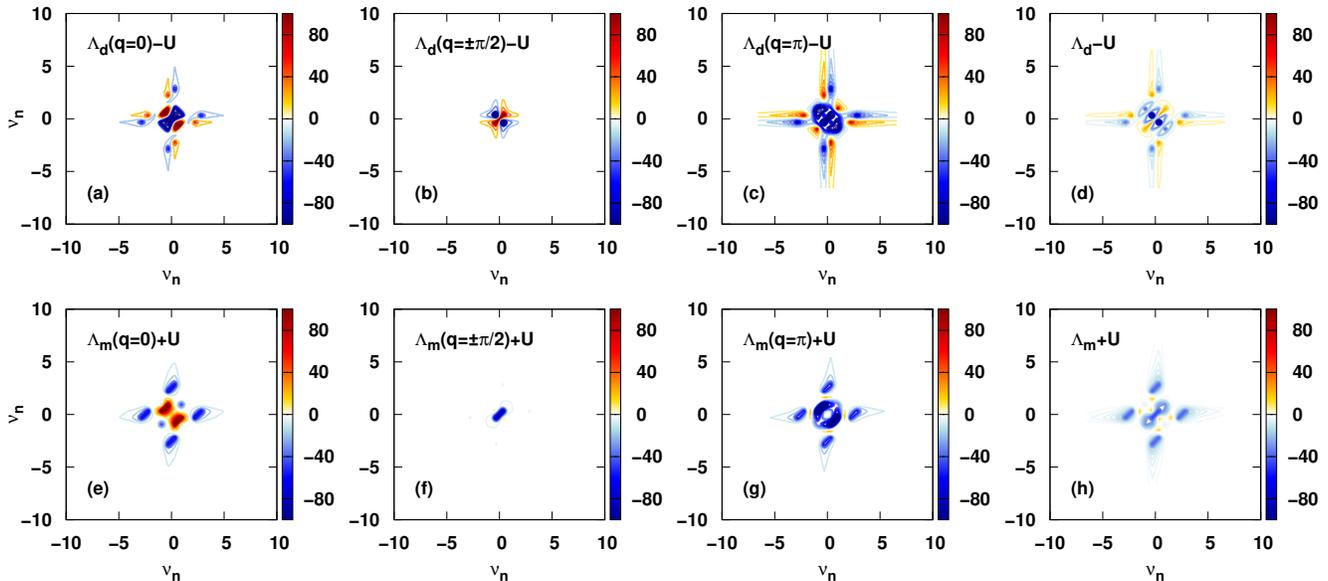}
\caption{(Color online) Exact fully irreducible vertex in the (particle-hole) density and magnetic channels 
with respect to the static asymptotics, i.e.: $\Lambda_d\!-\!U$ (upper row) and $\Lambda_m\!+\!U$ (lower row), 
as a function of the two fermionic frequencies $\nu_n$ and $\nu_{n'}$, for bosonic frequency $\omega=0$. 
The $q$-resolved vertex $\Lambda(q)$ (panel a, b, c, e, f, and g) corresponds to the fully irreducible vertex 
averaged over $k$ and $k'$, while the local $\Lambda$ (panel d and h) is averaged over $q$ as well.   
In addition to the non-trivial momentum structure of  $\Lambda(q)$, neglected within the parquet D$\Gamma$A, 
it is worth noting that the complex frequency structure of the local $\Lambda$ is not captured from the DMFT vertex  
(cfr. Fig. \ \ref{fig:fir_nc4U2T0.1}). This suggests that a full self-consistency at the two-particle level, 
via a corresponding redefinition of the AIM, might improve the present D$\Gamma$A results. 
Parameters: $N\!=\!4$, $U\!=\!2t$, and $T\!=\!0.1t$. }
\label{fig:firq_nc4U2T0.1}
\end{figure*} 

We discuss the effect of local and non-local correlations on the Green's function as we already did for the $N\!=\!6$ sites ring.  
Both in DMFT and D$\Gamma$A, a sizeable value of $G_{ii}(\beta/2)$ indicates a metallic spectral function, 
while in the exact solution this quantity is strongly suppressed, revealing an insulating nature. 
In this respect, we note that, even in the insulating state, a value of $G_{ii}(\beta/2)\!=\!0$ can only be achieved at $T\!=\!0$, 
while here we observe a finite value due to the average over an energy window 
due to the broadening of the Fermi distribution at finite temperature. 
The combined information of a sizeable value of $G_{ii}(\beta/2)$ and the large scattering rate $\gamma_k$ 
at the Fermi surface (i.e., $k\!=\!\pi/2$) in the corresponding self-energy in Fig.\ \ref{fig:siwk_nc4U2T0.1-parquet} 
suggest the presence of a local minimum in the spectral function at the Fermi level (pseudo gap). 
Hence, we can conclude that the D$\Gamma$A, in its full parquet-based implementation, 
yields a quantitative improvement over the DMFT description, 
however, the non-local correlations stemming from the 2PI local vertex of DMFT are not yet strong enough 
to completely open a well-defined gap in the spectral function, which is instead present in the exact solution. 

A deeper understanding of the above results can be obtained by the analysis    
of the frequency and momentum structure of the 2PI vertex.  
Let us first discuss the local 2PI vertex, shown in Fig.\ \ref{fig:fir_nc4U2T0.1}. 
The most striking feature of the vertex of the $N\!=\!4$ sites ring is the strongly enhanced low-frequency structure 
which now exhibits strong deviations from the bare interaction $U\!=\!2t$. 
In fact, the vertex corrections are orders of magnitude larger than for the $N\!=\!6$ insulating ring, 
and the low-frequency structure is also more complex. 
In particular, one can observe additional negative ''spots'' (of highest intensity) 
which are generated by the change of sign of several eigenvalues of the generalized local susceptibility.\cite{Schaefer2013} 
This low-frequency structure of the local 2PI vertex is responsible for a $k$-selective enhancement 
of the D$\Gamma$A self-energy over the one obtained within the PA. \\
The direct numerical evaluation of the exact 2PI vertex, shown in Fig.\ \ref{fig:firq_nc4U2T0.1}, 
allows to understand the role of its momentum structure. 
The $q$-resolved exact fully irreducible vertex $\Lambda(q)\!=\! \frac{1}{N_k^2}\sum_{kk'} \Lambda_{kk'q}$ 
shows that $\Lambda(q)$ displays a change in both sign and magnitude for different values of $q$ 
(the vertex is identical at $q\!=\!\pm\pi/2$ due to symmetry reasons). 
Such a large frequency and momentum dependence of the exact 2PI vertex can be possibly interpreted 
in terms of a proximity to a non-perturbative instability of the Bethe-Salpeter equations, 
such as those already reported for the Hubbard and Falikov-Kimball models.\cite{Schaefer2013,Janis2014,note_divergence} 
The strong momentum dependence of the fully irreducible vertex is certainly one of the reasons of the failure 
of the present D$\Gamma$A calculation to open a spectral gap at the Fermi level of the $N\!=\!4$ sites ring. 
However, an important piece of information is also enclosed in the exact local vertex $\Lambda\!=\!\frac{1}{N_q}\sum_{q} \Lambda(q)$. 
As shown in Fig.\ \ref{fig:firq_nc4U2T0.1}, the exact local $\Lambda$ displays a complex frequency structure, 
which is not fully captured by the local $\Lambda$ of DMFT (cfr. with Fig. \ \ref{fig:fir_nc4U2T0.1}). 
This suggests that, in this case, DMFT does not provide a good description of the two-particle local physics of the system. 
For this reason, performing a full self-consistency at the two-particle level, 
i.e., updating the local $\Lambda$ including the effect of non-local correlations, 
is expected to lead to improvements over the present D$\Gamma$A results. 
This idea is also supported by the calculations performed within the ladder approximation of the D$\Gamma$A, 
which we discuss in Sec.\ \ref{sec:ladder} in comparison with the parquet D$\Gamma$A. 
We anticipate that the divergent self-energy for $k\!=\!\pi/2$ is  already well approximated 
by the ladder D$\Gamma$A with Moriya corrections 
(which, however might fail in other regions of the DBZ, see also the related discussion in Sec.\ \ref{sec:ladder}).  
The results of the ladder approximation, shown in Fig. \ref{fig:siwk_nc4U2T0.1-ladder}, 
suggest that it is not the momentum structure of the 2PI to control the (large) self-energy, 
but rather the enhanced scattering induced by a strongly renormalized local vertex.  
Hence, a similar picture might also be obtained within a fully self-consistent parquet D$\Gamma$A scheme, 
where the local 2PI vertex will be further enhanced with respect to DMFT by non-local correlations. 

\begin{figure}[b!]
\includegraphics[width=0.47\textwidth, angle=0]{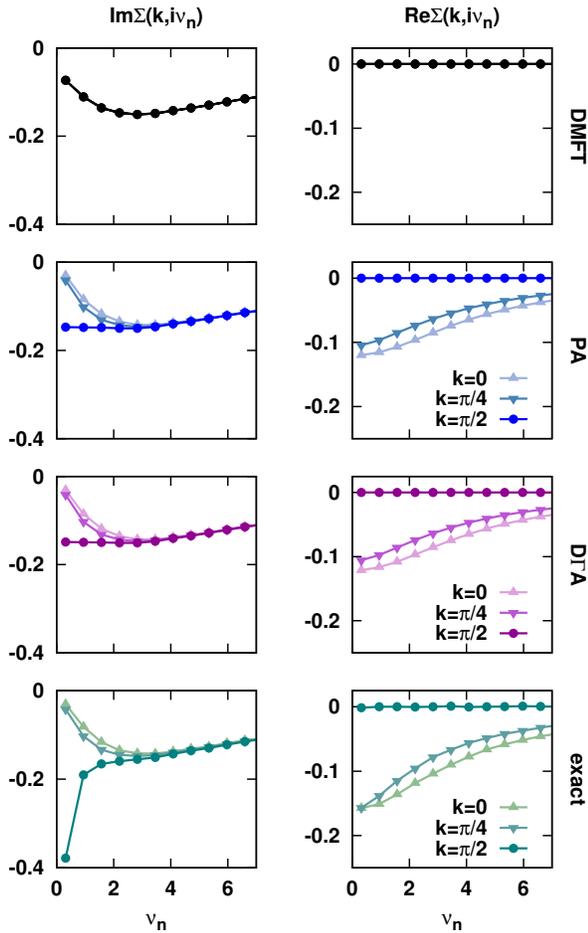}
\caption{(Color online) As in Fig.\ \ref{fig:siwk_nc6U2T0.1-parquet} but for the $N\!=\!8$ ring. 
In this case, including the full frequency dependence of $\Lambda$ results in negligible 
corrections to the self-energy, and the D$\Gamma$A results 
does not deviate appreciably from the one obtained within the static PA. }
\label{fig:siwk_nc8U2T0.1-parquet}
\end{figure}

\begin{figure}[t!]
\includegraphics[width=0.42\textwidth, angle=0]{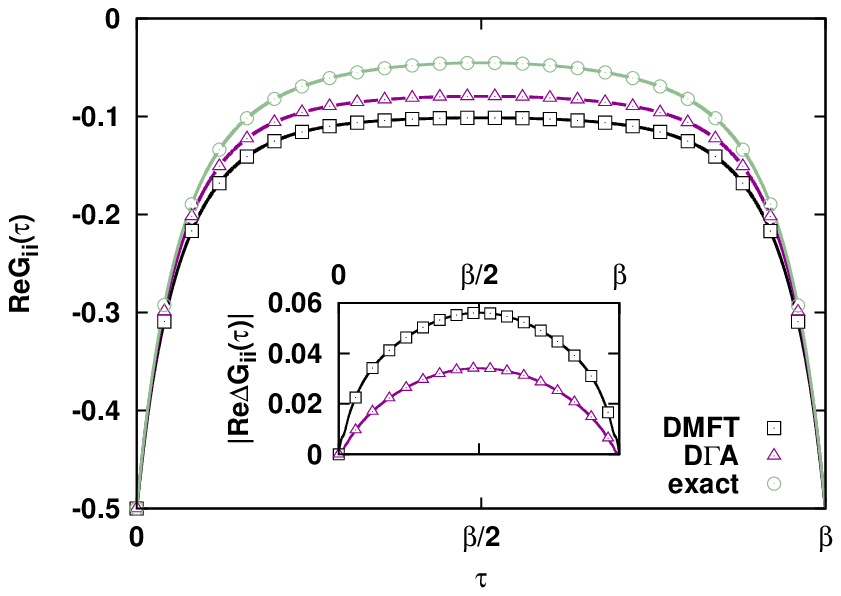}
\caption{(Color online) As in Fig.\ \ref{fig:gtau_nc6U2T0.1} but the $N\!=\!8$ ring.}
\label{fig:gtau_nc8U2T0.1}
\end{figure}

\begin{figure}[t!]
\includegraphics[width=0.5\textwidth, angle=0]{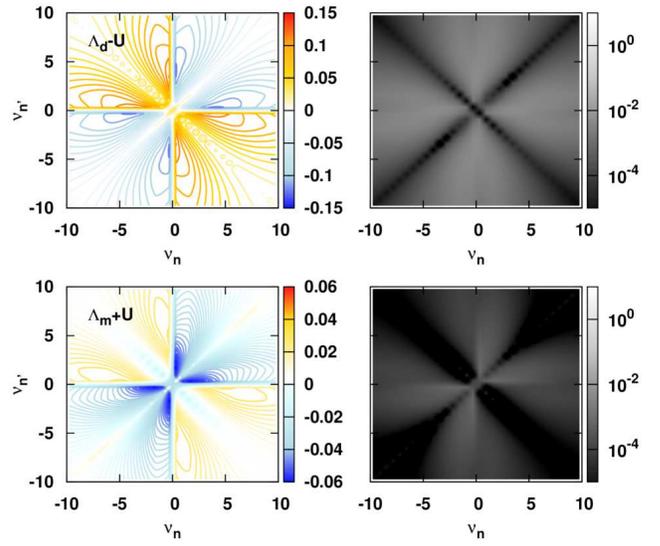}
\caption{(Color online) As in Fig.\ \ref{fig:fir_nc6U2T0.1} but for the $N\!=\!8$ ring.}
\label{fig:fir_nc8U2T0.1}
\end{figure}

We finally discuss the results for the $N\!=\!8$ sites ring, 
where the presence of additional structures in the non-interacting density of states, besides the 
(doubly degenerate) eigenstate at the Fermi level and the one at the band edge
lead to a somewhat different physical situation.  
Let us start discussing the $k$-resolved self-energy shown in Fig.\ \ref{fig:siwk_nc8U2T0.1-parquet}. 
As $N$ increases, the number of inequivalent $k$ points in the DBZ increases with respect to the previous cases. 
By symmetry it is sufficient to consider $k\!=\!0$, $k\!=\!\pi/4$, and $k\!=\!\pi /2$ (the latter at the Fermi surface).  
In this case, in contrast to the  $N\!=\!4$ sites ring, the DMFT self-energy shows a metallic bending, 
with a ($k$-independent) scattering rate $\gamma\!\approx\!0.1$. 
The comparison with the exact solution shows that the largest corrections with respect to DMFT 
are the enhanced scattering rate at the Fermi surface, $\gamma_{k\!=\!\pi /2}$, 
and the renormalization of the dispersion $\Delta_{k\!=\!0,\pi /4}={\rm Re}\Sigma(k, \imath\nu_n\rightarrow 0)$. 
The PA and the D$\Gamma$A give rise to similar non-local correlations, 
displaying a strong $k$-dependent scattering rate at the Fermi surface $\gamma_{k\!=\!\pi /2}\approx 0.3$. 
The large scattering rate reflects physically in the Green's function through a suppression of $G_{ii}(\beta/2)$, 
and hence of the low-energy spectral weight, with respect to DMFT, as shown in Fig.\ \ref{fig:gtau_nc8U2T0.1}. 
Although D$\Gamma$A provides an overall better description of the low-energy physics of the system 
with respect to DMFT, also in this case the parquet-based approximations 
fail to reproduce the divergent behavior of ${\rm Im}\Sigma(k\!=\!\pi /2,\imath\nu_n)$. 
Analogously to the case of the $N\!=\!4$ sites ring, the good agreement obtained within the ladder D$\Gamma$A 
for the self-energy at the Fermi surface, shown in Fig. \ref{fig:siwk_nc8U2T0.1-ladder}, 
points at the importance of the self-consistency at the two-particle level. 

As for the interpretation of the results, from the similarity of the PA and D$\Gamma$A results for the $N\!=\!8$ sites ring 
one would not expect a strong frequency dependence of the local 2PI vertex, 
as confirmed from the numerical data shown in Fig.\ \ref{fig:fir_nc8U2T0.1}. 
The 2PI vertex qualitatively resembles the one of the $N\!=\!6$ sites ring, 
with the difference that there is no suppression of the low-frequency structure. 
On the other hand, the difference between D$\Gamma$A and the exact solution 
might suggest an important momentum structure of the 2PI vertex.  
Unfortunately in this case a direct analysis is not feasible, due to the extremely high computational effort 
required to calculate the exact momentum-dependent two-particle vertex functions for the $N\!=\!8$ site ring. 
While a strong momentum dependence of the exact 2PI vertex is possible, 
also in this case the deviation observed between the parquet D$\Gamma$A and the exact solution 
might be induced by the poor approximation of the local physics of the system provided by the 2PI vertex of DMFT. 
This scenario, supported by the qualitatively correct behavior found within the Moriya corrected ladder approximation, 
suggests that the parquet D$\Gamma$A results might be further improved performing a fully self-consistent calculation.

\section{Relation to the ladder approximation}
\label{sec:ladder}
Hitherto, all the previous applications of the D$\Gamma$A scheme 
to bulk systems \cite{Toschi2007,Katanin2009,Rohringer2011,Schaefer-arXiv2014} 
were carried out within the ladder approximation only.  
This approximation is obtained by replacing the solution of the parquet equations in the flowchart of Fig. \ref{fig:flowchart}  
with a simpler calculation at the level of Bethe-Salpeter equations. 
Hence, in ladder D$\Gamma$A the non-local corrections to the local physics 
will be generated only in (a) selected channel(s). 
In practice, this corresponds to an essential simplification of the algorithm, 
because in ladder D$\Gamma$A only the corresponding irreducible vertex in the selected channel (e.g., spin) 
needs to be extracted from the AIM, and used to calculate the D$\Gamma$A self-energy via the Bethe-Salpeter equation. 

\begin{figure}[t!]
\includegraphics[width=0.47\textwidth, angle=0]{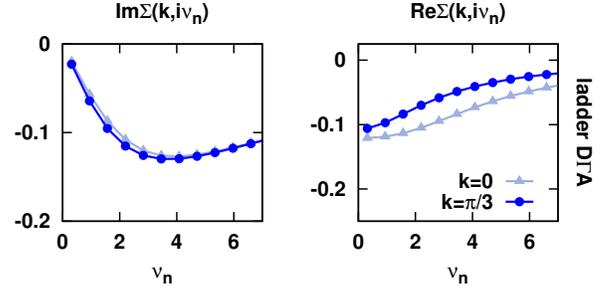}
\caption{(Color online) $k$-resolved  ladder D$\Gamma$A self-energy in Matsubara representation, 
for representative $k$-points in the DBZ. 
The ladder resummation was supplied with the Moriyasque correction to the spin propagator. 
Parameters: $N\!=\!6$, $U\!=\!2t$ and $T\!=\!0.1t$.}
\label{fig:siwk_nc6U2T0.1-ladder}
\end{figure} 

\begin{figure}[t!]
\includegraphics[width=0.47\textwidth, angle=0]{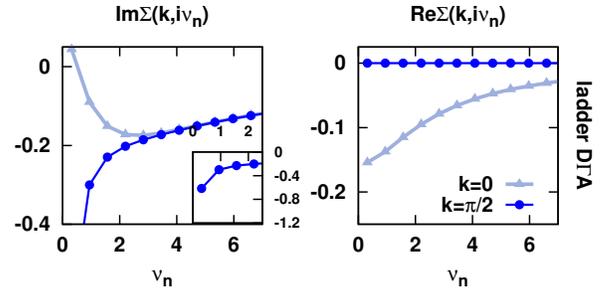}
\caption{(Color online) As in Fig.\ \ref{fig:siwk_nc6U2T0.1-ladder} but for the $N\!=\!4$ ring 
The non-causal self-energy for $k\!=\!0$ (grey dashed line) observed in this case is an extreme consequence 
of the breakdown of the ladder approximation far from the Fermi surface, as discussed in the text. 
The inset shows the low-energy tendency toward a divergency of the ladder D$\Gamma$A self-energy for $k\!=\!\pi/2$. }
\label{fig:siwk_nc4U2T0.1-ladder}
\end{figure} 

\begin{figure}[t!]
\includegraphics[width=0.47\textwidth, angle=0]{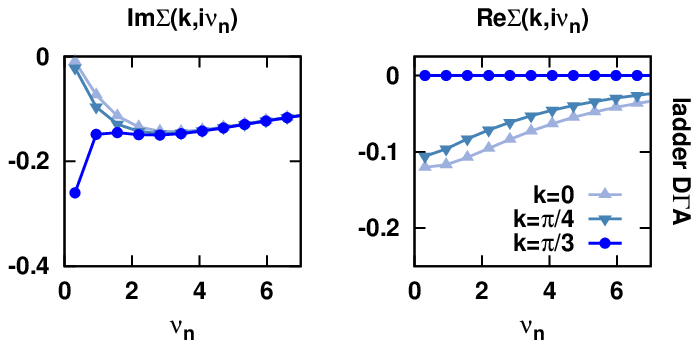}
\caption{(Color online) As in Fig.\ \ref{fig:siwk_nc6U2T0.1-ladder} but for the $N\!=\!8$ ring.}
\label{fig:siwk_nc8U2T0.1-ladder}
\end{figure}

\begin{figure*}
\includegraphics[width=0.8\textwidth, angle=0]{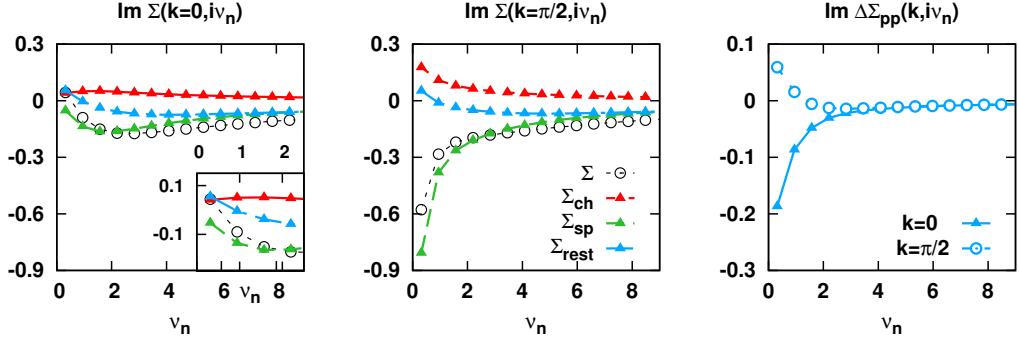}
\caption{[Left and middle panels] Fluctuation diagnostics of the ladder D$\Gamma$A self-energy, 
where ${\rm Im}\Sigma(k,\imath\nu_n)$ has been resolved in its contributions 
from the spin channel, the charge channel, and all the rest. 
Within the ladder approximation, at $k\!=\!\pi/2$ the contribution of the spin channel is dominant, 
while at $k\!=\!0$ all contributions are similar in magnitude. 
The inset shows the low-energy behavior of the different self-energy contributions at $k\!=\!0$.  
[Right panel] Non-local parquet D$\Gamma$A correction to the DMFT self-energy 
computed in the particle-particle scattering channel $\Delta\Sigma_{pp}$. 
Its strong $k$-dependence, neglected within the ladder approximation 
is at the origin of the causality violation of the ladder D$\Gamma$A self-energy at $k\!=\!0$ 
(cfr. Fig.\ \ref{fig:siwk_nc4U2T0.1-ladder}). 
Parameters: $N\!=\!4$, $U\!=\!2t$ and $T\!=\!0.1t$. }
\label{fig:fdiag}
\end{figure*}

The application of the ladder approximation is well justified in case the system displays predominating fluctuations 
in a given scattering channel, which is known {\sl a priori}, 
e.g., in the case of the antiferromagnetic instability in the 3D Hubbard model at half-filling.\cite{Rohringer2011} 
However, the significant numerical simplification of avoiding the solution of the direct and inverse parquet equations 
comes at the price of a more approximative approach, 
which is mitigated by performing the so-called Moriya-correction.\cite{Katanin2009,Held2008,RohringerPhD} 
This correction is done at the level of the propagator of the (e.g., spin) fluctuations, 
obtained from the generalized susceptibility by performing the sum 
over the fermionic Matsubara frequencies and corresponding momenta. 
In the Moriya scheme, a mass term $\lambda$ is added to the propagator, 
so that in the selected (spin) channel it reads 
\begin{equation}
\chi_s^{\omega}(q) ^{-1} \rightarrow  \chi_s^{\omega}(q)^{-1} + \lambda. 
\end{equation} 
As the mass is determined by imposing a condition over the local physics, 
the procedure mimics to a reasonable extent the effect of a full self-consistency of the algorithm, 
where also the local 2PI vertex would be renormalized by non-local spatial correlations. 

In the following, we discuss the results obtained within the ladder approximation of the D$\Gamma$A scheme 
for the self-energy of the $N\!=\!4, 6, 8$ sites rings. 
In the case of the $N\!=\!6$ sites ring, the ladder D$\Gamma$A self-energy, 
shown in Fig.\ \ref{fig:siwk_nc6U2T0.1-ladder}, is in good agreement with both the parquet D$\Gamma$A 
and the exact results (cfr. also Fig. \ref{fig:siwk_nc6U2T0.1-parquet} for a direct comparison). 
Slight deviations suggest that, although at half-filling the physics is expected to be dominated by spin fluctuations, 
in low-dimensions considering all the scattering channels (and their interplay) on the same footing, 
via the solution of the parquet equations, leads to quantitative corrections in this parameter regime. \\
The situation is different in the cases of the $N\!=\!4, 8$ sites rings. In particular, the ladder D$\Gamma$A self-energy, 
shown in Figs.\ \ref{fig:siwk_nc4U2T0.1-ladder} and \ref{fig:siwk_nc8U2T0.1-ladder}, 
is able to capture the large scattering rate at the Fermi surface $\gamma_{k\!=\!\pi/2}$ of the exact solution, 
improving over the parquet D$\Gamma$A results. 
This unexpected result is likely to be attributed to the ability of the Moriyasque corrections 
to mimic the self-consistency of the local (irreducible) vertex.\cite{Held2008,Katanin2009} 
This suggests that also the full parquet D$\Gamma$A might be able to reproduce 
the divergent trend of the self-energy at the Fermi surface 
with a better starting point for the local fully irreducible vertex than the one provided by DMFT. 
This would definitely be achieved by performing fully self-consistent D$\Gamma$A calculations.  

The ladder D$\Gamma$A calculations performed here also pointed out an important issue, 
i.e., the {\sl failure} of the ladder approximation far away from the Fermi surface. 
This is indicated for the $N\!=\!4$ sites ring by the non-causal self-energy 
obtained at the lowest Matsubara frequency for $k\!=\!0$. 
Several tests in this case ruled out the possibility that 
the non-analyticity of the self-energy is a physical artefact due to numerics 
(e.g., due to the finite frequency mesh). \\
By exploiting the (hitherto) unique possibility of having at disposal
both ladder- and parquet D$\Gamma$A self-energy and vertex results, we have performed a 
decomposition of the D$\Gamma$A self-energy, by separating the
contributions coming from the different channels, following a similar ``philosophy'' as 
in the recently introduced {\sl fluctuation diagnostics} for the electronic self-energy.\cite{Gunnarsson-arXiv:1411.6947} 
The assumption, under which a simplification of the parquet D$\Gamma$A algorithm 
to the ladder D$\Gamma$A is possible, is that the (k-dependent) non-local
corrections to the DMFT self-energy are dominated by the contribution of a specific channel (e.g., at half-filling, the spin channel). 
The fluctuation diagnostics of the ladder D$\Gamma$A self-energy, shown in the left and middle panels of Fig.\ \ref{fig:fdiag},   
demonstrate that this is indeed the case for the calculations of the self-energy at the Fermi level ($k\!=\!\pi/2$). 
On the other hand, we also see that, in the case of the $N\!=\!4$ sites ring,
the ladder assumption does not apply any longer far from the Fermi surface. 
In fact, as shown in Fig.\ \ref{fig:fdiag}, at $k\!=\!0$ the contribution of the spin-channel to the D$\Gamma$A self-energy 
is strongly reduced with respect to the $k\!=\!\pi/2$, becoming comparable with the contributions of the other channels. 
This means that the error introduced by ladder assumptions might become 
even larger than the value of ${\rm Im}\Sigma(k)$ itself, which is often strongly 
reduced by non-local correlation far from the Fermi surface. 
It is important to emphasize that the overall trend of a strong reduction of ${\rm Im}\Sigma(k\!=\!0)$ 
due to non-local correlation, which is also visible in the exact results, 
is correctly captured even by the ladder D$\Gamma$A calculations. 
However, quantitatively, the breakdown of the ladder assumption for this $k$-point 
leads to a large relative error on ${\rm Im}\Sigma(k\!=\!0)$, 
and eventually to an analyticity violation, preventing the applicability of the ladder D$\Gamma$A for this $k$-point. 
Our explanation is numerically supported by the comparison 
with the corresponding decomposition of the full parquet D$\Gamma$A self-energy. 
Specifically, in the right panel of Fig.\ \ref{fig:fdiag} we show the momentum dependence 
of the ``secondary''(particle-particle) channel contribution to ${\rm Im}\Sigma(k,\imath\nu_n)$. 
The D$\Gamma$A correction $\Delta\Sigma_{pp}$ is obtained by replacing 
$F^{\uparrow\downarrow} \rightarrow \Phi_{pp} - \Phi_{pp}^{\rm DMFT}$ in the equation of motion (\ref{eq:DSE}). 
At $k\!=\!0$, the correction with respect to DMFT (neglected in ladder D$\Gamma$A) is actually 
of the same order, if not {\sl larger}, than the contribution of the ``dominant'' channel 
and/or of the overall value of ${\rm Im}\Sigma(k\!=\!0)$, shown in the left panel of Fig.\ \ref{fig:fdiag}. 
Although the deterioration of accuracy of the ladder approximation far from the Fermi surface 
may be expected as a general trend, the error introduced is often not significant. 
For instance, in the parameter regime considered for the $N\!=\!8$ sites ring, causality is preserved. 
As the particle-hole channel is dominant in the vicinity of the Fermi level, 
neglecting the particle-particle channel is justified for this (most relevant) part of the DBZ, 
and the ladder approximation can still be employed.

\section{Summary and Conclusions}
\label{sec:SEC}
In this paper we have presented a numerical study of correlated Hubbard nano-rings, 
employing the parquet D$\Gamma$A.  
This algorithm corresponds to the actual realization of the original  D$\Gamma$A idea, 
in which the local (DMFT) assumption is made {\sl only} at the level of the 2PI local vertex of the theory, 
while non-local correlations beyond DMFT are computed 
simultaneously in {\sl all} channels by solving the corresponding parquet equations. 
This represents a methodological improvement over the ladder D$\Gamma$A algorithms used hitherto, 
where additional approximations (e.g., restriction to a given subset of ladder diagrams) were performed. 
The overall numerical effort of a full D$\Gamma$A calculation is clearly larger than  in the case of ladder approximations, 
but the numerical workload is still manageable in 1D and 2D.\cite{Yang2009,Tam2013} 

Specifically, we have shown results for correlated Hubbard rings of different sizes $N\!=\!4, 6, 8$, 
and performed a systematic comparison of the parquet D$\Gamma$A 
against DMFT, PA, ladder D$\Gamma$A and the exact QMC solution of the problem. 
We achieved a twofold goal: i) on the methodological side, 
we could test the accuracy of the D$\Gamma$A approximation for quasi-1D systems, 
which is arguably most challenging regime for the locality assumption of $\Gamma_{\rm irr}$; 
ii) on the physical side, we could understand the different roles played by
local and non-local correlations in determining the spectral properties of the systems considered. 
Our numerical calculations show that, for semiconducting nanostructures, 
the parquet D$\Gamma$A quantitatively reproduces the exact many-body solution of the system, 
improving over the corresponding local description of DMFT. 
Instead, if the bare dispersion displays a (doubly degenerate) peak at the chemical potential, 
the 2PI vertex acquires a non-trivial frequency and momentum dependence. 
As the local $\Lambda$ of DMFT captures, to some extent, the dynamical structure of the 2PI vertex, 
the parquet D$\Gamma$A, although not perfect, provides a better qualitative description than those of DMFT and the static PA. 
Updating the 2PI vertex, i.e., performing fully self-consistent D$\Gamma$A calculations, 
is expected to further improve the present results. However, this lies beyond the scope of the present work. 

In conclusion, we have shown how the parquet D$\Gamma$A algorithm can be implemented and applied with success 
to analyze the physics of correlated Hubbard nano-rings. 
Exploiting the recent improvements in the calculation of the local vertex of DMFT\cite{Rohringer2012,Hafermann2014} 
and in the numerical solution of the parquet equations,\cite{Yang2009,Tam2013}  
our results, which are obtained in one of the most difficult regimes for the D$\Gamma$A, 
pave the path for a more accurate theoretical treatment of strongly correlated electron systems.

\begin{acknowledgements}
We thank M.~Capone, P.~Hansmann, A.~Katanin, V.~Meden, and C.~Taranto for useful discussions. 
We also thank O.~Gunnarsson for making his Hirsch-Fye QMC program available, 
as well as H.~Fotso, K.~M.~Tam, and M.~Jarrell for discussions on the parquet solver, 
available at \url{http://www.phys.lsu.edu/~syang/parquet/}, which we adapted to perform the D$\Gamma$A calculations.  
We acknowledge financial support from the Austrian Science Fund (FWF) through I-610-N16 (AV, GR, AT), 
the FWF  Doctoral School ''Building Solids for Functions'' (TS) and SFB ViCoM F41 (SA), 
the Deutsche Forschungsgemeinschaft through FOR 1346 (GS),  ZUK 63 (SA), SFB/TRR 21 (SA), 
and the European Research Council under the European Union's Seventh Framework Program 
(FP/2007-2013)/ERC through grant agreement n.~306447 (PT, KH). 
The numerical calculations have been performed on the Vienna Scientific Cluster (VSC). 
\end{acknowledgements}

\appendix*

\section{Computational Details}
In the following we provide the technical details for each step of our calculations. 
\indent Both the nano-DMFT and exact calculations were performed employing a Hirsch-Fye QMC impurity solver. 
This allowed us to compare the results obtained on the same footing, 
i.e., with the same systematic discretization error $\Delta\tau$, 
as already done in previous studies. \cite{Valli2010,Valli2012} 
Here we compared QMC calculations performed for values of $\Delta\tau\!=\!0.1\overline{6}$ and $\Delta\tau\!=\!0.08\overline{3}$. 
Although no systematic extrapolation to $\Delta\tau \!\rightarrow\! 0$ \cite{Bluemer2007} has been done, 
we verified that the DMFT data are substantially converged in $\Delta\tau$ 
by a direct comparison against independent continuous-time QMC calculations. 
Also, we tested the exact solution against the exact diagonalization of the Hamiltonian. 
In both cases we obtained satisfactory agreement. \\
\indent In this work, we evaluated the local two-particle vertex functions of DMFT with an exact diagonalization (ED) impurity solver. 
To this end, we fit the hybridization function of the AIM corresponding to the converged DMFT(QMC) calculation 
with a discrete number of sites $N_s$ in order to obtain the corresponding Anderson parameters. 
We performed calculations with $N_s\!=\!5$ sites, i.e., $N_b\!=\!4$ bath sites and one impurity, 
which represents the current numerical limitation of ED for the calculation of two-particle Green's function, 
in its full frequency dependence, denoted as $G^2(\nu,\nu',\omega)$ in the notation of Ref.\ \onlinecite{Rohringer2012}.
ED calculations with (few) more bath sites might become feasible in the future, 
under specific conditions, e.g., by exploiting some additional symmetries of the AIM, 
and further optimizing the algorithm.\cite{pomerol}   \\
As for the present calculations, we have verified the accuracy of our numerical data for the vertex 
against continuous-time QMC calculations, and, for $N_s\!=\!5$, 
we have found very satisfying agreement at low (Matsubara) frequencies. 
At high frequencies, the ED data generally display a better asymptotic behaviour with respect to QMC data, 
despite the schemes proposed to overcome this issue. \cite{Gunnarsson2010,Boehnke2011,Hafermann2012} 
As the calculation of fully irreducible vertex functions, as well as the solution of the parquet equation involve several 
matrix inversions, the asymptotic behaviour of vertex functions plays a crucial role 
for the stability of the algorithm and the numerical accuracy of the final D$\Gamma$A results. 
This motivates our choice of ED as (two-particle) impurity solver for the present study, 
while not excluding different choices for the future. 
The calculation of the two-particle Green's function from its Lehmann representation \cite{Toschi2007,Rohringer2012} 
scales polynomially with the number of frequencies $N_f$ as {\cal O}($N_f^3$), 
and exponentially with the number of bath sites. 
In our calculations we obtained the full vertex $F$ within ED up to $N_f\!=\!320$ Matsubara frequencies.  
In each case, $N_f$ have been chosen in order to get, at the temperature considered, 
a fully irreducible vertex $\Lambda$ displaying a smooth asymptotic behavior towards its static value at high-frequency. 
Typical calculations of the two-particle Green's function performed in this work require about $50.000$ CPU-h. 
\indent The numerical solution of the parquet equations\cite{Yang2009,Tam2013} scales as {\cal O}($N_t^4$), 
where $N_t\!=\!N\!\times\!N_f$. For the systems considered here, i.e., for $N\!=\!4, 6, 8$, 
we performed calculations on a mesh of $N_f \!=\! 100$ Matsubara frequencies. 
The input of the parquet solver is the local fully irreducible vertex of DMFT, 
calculated as discussed above on the corresponding frequency mesh. 
Typical self-consistent parquet calculations performed in this work require about $1.000$ CPU-h, 
depending on both the value of $N_t$ and of the convergency threshold. \\
The relatively limited number of frequencies available results in numerical issues, 
e.g., the internal sums over the frequencies in the Dyson-Schwinger equation of motion 
may fail to yield the correct $1/\nu$ high-frequency behavior for the imaginary part of the $D\Gamma$A self-energy, 
and also decreases the overall accuracy of the D$\Gamma$A calculation. 
Since the high-energy tail of the self-energy is entirely determined by the bare interaction $U$, 
we overcome the issue performing a correction using the tail of the DMFT self-energy. 
To this end, we compute a local self-energy $\Sigma^{*}$ solving the local parquet equations, 
using as an input the local $\Lambda$ and ${\cal G}_{0}$ of DMFT. 
We correct the self-energy of D$\Gamma$A by subtracting $\Sigma^{*}$ and adding the $\Sigma$ of DMFT. 
Calculations for larger systems or with more frequencies than the one presented here are possible, 
although network communication and memory storage, rather than the computational time, become the main bottlenecks.


\begin{thebibliography}{99}
\bibitem{note_siwk} Due to the particle-hole symmetry 
and to the degeneracy of the non-interacting eigenstates, the other components of the $k$-resolved self-energy 
can display, at most, a sign change with respect to the ones shown here.
\end{thebibliography}
\end{document}